\documentstyle[preprint,aps,eqsecnum]{revtex}
\begin{document}
\title {Quantum point contact in a compressible Quantum Hall liquid}

\author {D.V. Khveshchenko}

\address {NORDITA, Blegdamsvej 17, Copenhagen DK-2100, Denmark}

\maketitle

\begin{abstract}
\noindent
We consider electron transport through a  
quantum point contact between compressible Quantum Hall liquids and derive the contact's 
impedance function for both diffusive and ballistic 
regimes of the bulk electron motion. In either regime, the tunneling current deviates from   
a power-law behavior $I\sim V^{1/\nu}$ expected on the basis of
the single-mode chiral Luttinger liquid phenomenology.

\end{abstract}
\pagebreak

Over the past few years a great deal of attention was paid to the  
transport properties of incompressible   
Fractional Quantum Hall Effect (FQHE) edge states, which, following the seminal work by 
Wen \cite{1}, have been customarily viewed as comprised of a 
finite number of chiral one-dimensional (1D) edge modes. 
However, as a consequence of the intrinsic discontinuity in the FQHE ground state and 
excitations' properties as a function of the filling factor $\nu$, the  
 phenomenological description of Ref.\cite{1} 
implied that the number of chiral 1D modes can vary drastically with  
$\nu$.

According to Ref.\cite{1}, the simplest single-mode chiral Luttinger liquid ($\chi$LL) 
description can only hold for the Laughlin filling fractions $\nu=1/(2p+1)$. 
A power-law edge electron's density-of-states (DOS) $\rho(\omega)\sim 
\omega^{g-1}$ with $g=1/\nu$  
as well as other distinct hallmarks of the $\chi$LL behavior 
were predicted to manifest themselves in tunneling experiments. 

In practical terms, this would limit the applicability of the 
phenomenological theory envisioned in Ref.\cite{1} to the edge of the $\nu=1/3$ FQHE,
  since, according to the experiment,  
Wigner crystallization can set in already at $\nu=1/5$.

Until recently, an extension of the phenomenological single-mode theory of Ref.\cite{1} 
was only proposed for the incompressible Jain fractions $\nu={n/(2pn+1)}$ \cite{2}.  
In Ref.\cite{2} these FQHE states were described in terms of $n$ edge modes, which either 
all propagate in the same direction 
(the chiral case $n>0$) or one of them goes in the direction opposite to that of the other 
$n-1$ modes 
$(n<0)$. 

The latter case appears to be much more complicated. In particular,
the authors of Ref.\cite{2} pointed out that in order to have the two-terminal 
conductance 
equal to its expected universal value given by the bulk Hall conductivity
$\sigma_{xy}=\nu e^2/h$ one has to include  
disorder, which then provides momentum-non-conserving scattering 
between different modes and facilitates their equilibration. Having equilibrated,
all $n$ edge modes were found to contribute comparably to the universal DOS 
exponent $g=1+|2p+1/n| - 1/|n|$.   

Further attempts to find a microscopic justification of this phenomenological
picture showed that even for the simplest 
fractions it can only hold
for an atomically sharp edge whose width $W$ is smaller than the magnetic length $l_B$. 
On the contrary, the theory of a smooth edge   
yields the DOS exponent $g(W)$ which grows with $W$, so that   
 $\rho(\omega)$ undergoes 
a crossover to the gapful bulk behavior  $\sim \exp(-\Delta/|\omega|)$ as $W$ tends to 
infinity \cite{3}.

Compared to the case of FQHE, the edge-related physics of compressible Quantum Hall (QH)  
liquids received much less attention, largely because of the lack of an appropriate 
theoretical description.
Indeed, the edge states theory of Ref.\cite{1} 
can not be naively extended onto a gapless "composite fermion (CF) metal" 
which is believed to form at even denominator fractions $\nu\sim 1/2p$ \cite{15}.

Alternately, in \cite{5} the edge tunneling problem   
was considered for both compressible and incompressible states in the unifying 
framework of the systematic CF 
theory of Ref.\cite{15}. In their analysis, the authors of Ref.\cite{5} restricted 
themselves onto the diffusive regime 
of the bulk dynamics of CFs governed by short-range (screened
Coulomb) interactions. 

According to Ref.\cite{5}, the   
tunneling DOS exhibits a power-law  
behavior with the exponent given in terms of the physical bulk conductivities: 
$g=1+{h\over e^2}|\sigma_{xy}|^{-1}(1-|1-2p\sigma_{xy}|)+O(\sigma_{xx})\approx min 
(2\nu^{-1}+1-2p, 1+2p)$
in the range of filling fractions ${1/(2p+1)} < \nu < {1/(2p-1)}$, provided
that $\sigma_{xx}<<\sigma_{xy}$.
In the limit $\sigma_{xx}\rightarrow 0$ at 
$T\rightarrow 0$ this result agrees with the predictions
of the phenomenological theory of Ref.\cite{2} at all Jain fractions.  

However, these theoretical results were called into question after a direct 
tunneling experiment by Grayson {\it et al} 
revealed a power-law behavior with the exponent $g\approx 1/\nu$ \cite{6}.
A systematic (of order $10 {\%}$) deviation  towards smaller values of $g$ 
observed in a wide range of $\nu$ was attributed in \cite{6}
 to a $\sim 20{\%}$ pile-up of electron density near the edge caused by Coulomb 
interactions.
  
Although, naively, the exponent $g=1/\nu$ can be readily derived from the simplest 
single-mode $\chi$LL theory of Ref.\cite{1},
the findings of Ref.\cite{6} did pose a challenge, since no 
such a description was expected to exist for generic filling fractions, 
especially for the compressible ones. 

It was this puzzle that inspired the latest theoretical
developments \cite{17}, according to which all Jain fractions allow for an alternate 
description 
in terms of a single propagating edge mode that carries physical charge plus a pair of 
non-propagating 
fields required to guarantee a compliance with Fermi statistics. Unlike the original theory 
of Refs.\cite{1,2}, 
no propagating neutral modes appear in its refined version of Ref.\cite{17} (see also 
\cite{18} for a related discussion).

This new phenomenology corroborates the conclusions drawn from previous microscopic 
analyses of realistic edges governed by Coulomb 
interactions \cite{16} which provided a host of compelling evidence that tunneling electrons 
can hardly couple to any neutral edge modes (if any), but only to the charge-carrying one 
(edge magnetoplasmon).

In the present paper, we complement the previous efforts
and address the problem of tunneling to the edge of a compressible QH liquid.

According to the picture proposed  in \cite{7}, tunneling from a metallic lead to the QH 
edge occurs via 
isolated impurities which can be formally thought of as atomic-size
quantum point contacts (QPCs). Furthermore, by generalizing the argument   
of Ref.\cite{7} to the interacting case, one can show that, as long as interactions 
in the lead do not alter its normal Fermi liquid behavior characterized by
the $\sim 1/t$ decay of the electron Green function $G(t,{\bf 0})$, one can formally 
map the only lead's conduction channel that couples to the QH edge onto a chiral 1D Fermi 
liquid. 
In turn, the latter is formally equivalent to the dissipationless $\nu=1$ QH liquid. 
Thus one can put edge tunneling into the context of the     
problem of charge transport through an atomic-size QPC between
two (in general, different) QH liquids. 

The previously studied problem of a QPC between 
two FQHE states, each having a well-defined edge 
described by a single chiral boson $\Phi_{R,L}(x,t)$,
reduces to a local (zero-dimensional) imaginary-time ($0<t<\beta=1/T$) 
action for the out-of-phase combination $\Phi(t)=\Phi_R({0},t)-\Phi_L(0,t)$ \cite{8}:     
$$ S= 
{1\over 4\pi\beta}\sum_{\omega} {|\omega|\over g(\omega)}|\Phi(\omega)|^2 +\Gamma 
\int^{\beta}_{0} dt\cos \Phi(t) 
\eqno(1)$$  
where $\Gamma$ is a bare electron tunneling amplitude. In the inhomogeneous case 
$\nu_L\neq\nu_R$ 
the coupling function $g(\omega)$ is given  by
the mean of the two functions characterizing the
FQHE states on either side of the QPC located at ${\bf r}=0$ \cite{7}:
$$g(\omega)={1\over 2}(g_L(\omega)+g_R(\omega))  \eqno(2)$$ 
In the absence of long-range Coulomb interactions each coupling function is merely a 
constant
$g_{L,R}(\omega)=1/\nu_{L,R}$, and, both, the current-voltage characteristic $I(V)$
and the linear conductance $G(T)=dI/dV$ can be found from the exact solution of the related
1D boundary sine-Gordon model \cite{9}.  

At voltage biases or temperatures large compared to an intrinsic crossover energy scale
$eV_0\propto \Gamma^{\nu/(\nu-1)}$ the two-terminal conductance 
approaches its peak value $G=\sigma_{xy}$
corresponding to the perfect transmission ($\Gamma\rightarrow 0$), while
at low $V$ or $T$ the  
QPC goes to the pinch-off regime, and the residual (non-ohmic) conductance is solely due to 
electron 
tunneling through the potential barrier. 

In the experiment on an 
electrostatically defined constriction in a uniform $\nu=1/3$ FQHE state \cite{10}, 
the measured $G(T)$ was 
found to deviate from the result of the exact calculation of Ref.\cite{9} at low $T$.
Conceivably, this departure calls for a need to include long-range Coulomb interactions
between physically separated edges, in which case the coupling function 
becomes frequency dependent \cite{11}.

We are now going to consider a QPC between compressible QH
liquids and derive the local Caldeira-Legett-type action (1), thereby showing that it may 
hold 
regardless of the existence of well-defined edge modes.

To this end, 
we employ a bulk version of the Independent Boson Model (IBM)
formulated in terms of the local charge densities $\rho(t,{\bf r})$. 
The earlier, somewhat heuristic, applications of this method to the analysis of 
the bulk properties of QH systems \cite{13} received their firm microscopic justification in 
\cite{14}.
 
The IBM approach is commonly viewed as an adequate description of sudden 
shake-up processes, such as an X-ray emission or absorption, 
which create a large number of low-energy bosonic collective modes 
decaying on a long time scale and, therefore, dominating in the total action 
$S(t)$ or 
in transient amplitudes $<out|e^{-iHt}|in>\sim \exp(-S(t))$ characterizing 
the process \cite{12}.

Here we deal with another example of such a process, single-electron tunneling, 
that causes a sudden local charge perturbation which then slowly 
relaxes on a time scale $t\sim 1/eV$ set by the applied bias and spreads into the 
bulk over a distance $L\sim t^{\alpha}$ where $\alpha$ is either $1/2$ or $1$, 
depending on the regime (either diffusive or ballistic) of the bulk electron motion.

It is the insight gained in Refs.\cite{17,16} which makes us believe 
that there is no other propagating ("neutral") collective mode that can be excited upon
electron tunneling and then needs to be taken into account. 

For low biases $V$ the length scale $L\sim V^{-\alpha}$ is by far larger than the size of 
the QPC,
which allows one to neglect the detailed structure of the QPC and to model it
as a pinhole in a screen extended along the line $x=0$.

In the IBM approach, the necessary semi-microscopic input is provided in the form 
of the bulk density correlation function 
$\chi (\omega, {\bf q})=<\rho(\omega,{\bf q})\rho(-\omega,-{\bf q})>$.
In the case of metal-like CF states, different regimes (ballistic versus diffusive) as well 
as different pairwise 
interaction potentials (Coulomb versus screened) can be all incorporated into one expression 
\cite{15}:
$$\chi^{-1}(\omega,{\bf q})={|\omega|\over \sigma_{xx}({\bf q}) {\bf q}^2}+U_{\bf 
q}    \eqno(3)$$
For the sake of simplicity, hereafter we put $T=0$ and assume that both 
crossovers (from ballistic to diffusive bulk dynamics and from Coulomb to short-range 
interactions due to screening by the 
metallic parts of the Hall device) occur in the same range of biases 
$eV\sim \tau^{-1}$ given by the bulk CF impurity scattering rate.

The Lagrangian of the corresponding IBM model then reads as
$$L=\sum_{\alpha =L,R}\biggl(
\psi^{\dagger}_{\alpha}(-\partial_t+\mu_{\alpha})\psi_{\alpha}
+\sum_{\bf q}({1\over 2}\rho_{\alpha}\chi^{-1}_{\alpha}\rho_{\alpha}
+ \rho_{\alpha}U_{\bf q} \psi^{\dagger}_{\alpha}\psi_{\alpha})\biggr) 
+\Gamma (\psi^{\dagger}_{L}\psi_{R}+ \psi^{\dagger}_{R}\psi_{L}) \eqno(4)$$
where $\mu_R-\mu_L = eV$.
In the spirit of Refs.\cite{13,14}, in Eq.4 we neglect 
electron's recoil in processes of emission and absorption of the bulk charge 
density modes.

First, we consider the case of short-range (screened) interactions $(U_{\bf q}=U_0)$ 
in the diffusive regime ($\sigma_{xx}({\bf q})=\sigma_{xx}$).
By applying the linked cluster expansion \cite{12} 
to the Lagrangian (4) one 
can establish its equivalence (in the sense of amplitudes of all 
electron number-conserving transitions, see Ref.\cite{14} for a related procedure)
to a purely bosonic theory for
coupled L- and R- density modes 
$$L=\sum_{\alpha=L,R}(
{1\over 2}\rho_{\alpha}\chi^{-1}_\alpha\rho_{\alpha}+\mu_\alpha \rho_{\alpha})+
\Gamma \cos \biggl(U_0\int^t_0 dt^\prime(\rho_{R}(t, {\bf 0})-
\rho_{L}(t, {\bf 0}))\biggr)    \eqno(5)$$
Next, we integrate over gaussian fluctuations of local densities
 $\rho_{\alpha}(t,{\bf r})$ at ${\bf r}\neq 0$ and single out the out-of-phase 
combination $\rho_{-}(t,{\bf 0})=\rho_R(t,{\bf 0})-\rho_L(t,{\bf 0})$
which is the only one affected by the tunneling term. As a result, we arrive at 
the action 
(1) written in terms of the phase variable $\Phi(t) =U_0\int^t_0 
dt^{\prime}\rho_{-}
(t^{\prime}, {\bf 0})$ conjugated to the electron charge of the QPC. 

The coupling function $g(\omega)$  
resulting from the gaussian integration 
is given by the inverse kernel of the quadratic form appearing in Eq.5: 
$$g(\omega) = 
{U_0\over \pi}<{\bf 0}|(|\omega|+\nabla_{i}D_{ij}\nabla_{j})^{-1}|{\bf 0}>  
\eqno(6)$$
where $D_{ij}=U_0\sigma_{ij}$ is the tensor of bulk diffusion 
coefficients.

The differential operator in (6) has to be inverted by taking into account the 
"tilted" 
boundary condition 
for the current: $ J_x=-D_{xx}\nabla_x\rho - D_{xy}\nabla_y\rho = 0$ which is to 
be imposed 
everywhere
along the line $x=0$, except for the location of the QPC:
$$<{\bf r}|(|\omega|+\nabla_{i}D_{ij}\nabla_{j})^{-1}|{\bf 0}>=
\int {dk\over 2\pi} {e^{iky-|x|{\sqrt {k^2+|\omega|/D_{xx}}}}
\over D_{xx}{\sqrt {k^2+|\omega|/D_{xx}}}+iD_{xy}k},  \eqno(7)$$
the upper cutoff in the integral being $|k|\sim (v_F\tau)^{-1}$. Calculating the 
integral 
at $\omega <<\tau^{-1}$ and assuming that $\sigma_{xx}<<\sigma_{xy}$, 
we obtain   
the explicit form of the coupling function 
$$g(\omega)= {2e^2\over \pi h(\sigma^2_{xx}+\sigma^2_{xy})}
(\sigma_{xy}\tan^{-1}{\sigma_{xy}\over \sigma_{xx}}+{1\over 2}
\sigma_{xx}\ln {1\over |\omega|\tau}) \eqno(8)  $$
which has a meaning of an effective impedance of the electrodynamic 
environment
created by the bulk density modes.

The current through the QPC is given by the Kubo formula
$$I(V)= {e^2\over 2\pi h}\int^{\infty}_{0} dt e^{iVt} 
<[\partial_t\Phi(t), \partial_t\Phi(0)]>  
\eqno(9)$$
At low biases electron tunneling is weak, and 
the non-ohmic $I-V$ characteristic can be recovered from (9) in the second 
order in renormalized (and, therefore, energy-dependent)
tunneling amplitude $\Gamma(V)$: $dI/dV\sim |\Gamma(V)|^2$.
The latter obeys the renormalization group equation \cite{8}
$${d\ln \Gamma(\omega)\over \ln\omega}=g(\omega)-1 \eqno(10)$$
In the inhomogeneous case $\nu_L\neq\nu_R$,  
one can make Eq.10 more physically transparent by distinguishing  
between two separate tunneling amplitudes $\Gamma_{LR}(V)$ and $\Gamma_{RL}(V)$
which have the same bare value $\Gamma$ but can go apart upon renormalization. 
The second order $(\sim |\Gamma|^2)$ term turns out to be a product of the two,
which is consistent with the coupling function in Eq.10 being given by the combination rule 
(2).

For the experimental setup of Ref.\cite{6} where a compressible
QH liquid at $\nu_L=\nu$ is brought into a contact with a metallic lead  
described, according to Ref.\cite{7}, by $g_R=\nu_R^{-1}=1$, 
 Eq.10 yields 
$$I(V)\propto ({V\tau})^{{1/\nu}}
\exp\biggl(- 
{\sigma_{xx}h\over {2\pi e^2\nu^2}}|\ln V\tau|^2\biggr)    \eqno(11)$$
The logarithmic deviation of 
$g(\omega)$ from the value 
$1/\nu$ has the same origin as the diffusive exchange  
correction to the bulk tunneling DOS which is well-known in the zero field case. 
It is therefore starkly different
from the effect of logarithmic corrections to the FQHE edge mode's velocity 
due to long-range Coulomb interactions, as discussed in Refs.\cite{11}.

Our Eqs.8,11 lend further support to a physically motivated expectation that in the presence 
of 
disorder all observables must depend smoothly on the components of the conductivity tensor
$\sigma_{ij}(\nu)$. 
The experiment of Ref.\cite{6}
is suggestive of a possibility that, at least for some of the observables,
 it may be possible to extend these formulae "by continuity" onto 
incompressible FQHE states
by sending  $\sigma_{xx}$ to zero, in which limit
the $I-V$ characteristic becomes a pure power-law. In this way, one can make
a link to the single-mode description of the 1D edge states elaborating on the fact 
that in the limit $\sigma_{xx}\rightarrow 0$ the charge disturbance 
caused by tunneling electron does not propagate inside the bulk but only spreads 
along the edge in the direction set by the magnetic field 
(for arbitrary $\sigma_{xx}$ the charge spreading into the bulk relative to that 
along the edge scales as $L_x/L_y\sim \sigma_{xx}/\sigma_{xy}$, according to Eq.7).

Since the bulk resistivity of a typical CF state is low  
($\rho_{xx}\sim 10^{-1}-10^{-2} h/e^2$ which corresponds to $\tau^{-1}\sim 10^2\mu V$),  
the $I-V$ characteristic features an approximately power-law behavior $I\sim V^{1/\nu}$  
in the whole interval of applied biases $\tau^{-1} \exp (-\pi h/\nu\rho_{xx}e^2)< V < 
min\{V_0,\tau^{-1}\}$.

A substantial deviation from the exponent $1/\nu$ can only develop at very low biases where
 Eq.11 obtained under the assumption of a 
relative smallness of the logarithmic DOS correction has to be modified, too.

In Ref.\cite{6}, the data taken in the range of biases from  $10^0$ to 
$10^4\mu V$ indicate that the weak-to-strong coupling crossover occurs at $V_0\sim 10^3\mu 
V$, which
implies that the ballistic weak-coupling regime can also be probed experimentally.

Given that screening becomes less effective at higher energies, the ballistic regime is more
likely to be governed by the unscreened 
Coulomb interactions $U_{\bf q}=2\pi e^2/\epsilon_0q$. 

In the ballistic regime 
$q>(v_F\tau)^{-1}$ the bulk CF conductivity  
exhibits its peculiar linear momentum dependence: 
$\sigma_{xx}({\bf q})={e^2q\over (2p)^2hk_F}$ \cite{15}. 
As a result, the bulk charge density mode retains its diffusive spectrum
$\omega=iD{\bf q}^2$ with an effective diffusion coefficient 
$D={h\over 2\pi e^2}\sigma_{xx}({\bf q})U_{\bf q}= e^2l_B/4\epsilon_0p^{3/2}$.
 
Although in the immediate vicinity of the barrier the real-space density
correlation function $\chi(\omega, {\bf r}, {\bf r}^{\prime})$ ceases to be
translationally invariant and depends on both variables 
${x}\pm {x}^{\prime}$,  at distances
from the barrier satisfying the condition $L >> max({1/k_F, \epsilon_0/e^2m})$ its bulk-like 
behavior gets restored 
(see \cite{19} for a closely related discussion). Incidentally, we are only interested
in large length scales $L\sim (D/eV)^{1/2}$ 
which dominate in the tunneling action $S(t)$ for times $t\sim 1/eV$.
 
By using our definition of the effective $D$ and the mean field value of
the CF mass $m=k_F\epsilon_0/Ce^2$ with $C\approx 0.3$ \cite{15} we observe 
that for nearly all biases  smaller than the CF Fermi energy $E_F\sim 10^{3}-10^{4}\mu V$
 the above conditions 
are satisfied.

After inverting the corresponding 
diffusion operator, 
we calculate the matrix element (6) which now yields 
$$g(\omega)={e^2\over h}\int {dk\over \pi} 
\biggl({\epsilon_0\over e^2}{\sqrt {|\omega|(|\omega|+Dk^2)}}\ln^{-1} {{\sqrt 
{|\omega|+Dk^2}}+{\sqrt |\omega|}
\over {{\sqrt {|\omega|+Dk^2}}-{\sqrt |\omega|}}}+ik\sigma_{xy}\biggr)^{-1}
\eqno(12)$$
where the integral has the upper cutoff $|k|\sim 1/l_B$. 
Computing the integral (12) at frequencies $\tau^{-1}<< \omega <<E_F\sim 
e^2l_B/\epsilon_0$, we 
arrive at the expression  
$$g(\omega)={e^2\over \sigma_{xy}h}\biggl(1+{2\over |\ln\omega/E_F|}+
O({\ln |\ln \omega/E_F|\over |\ln \omega/E_F|^2})\biggr) 
 \eqno(13)$$ 
which can deviate from $1/\nu$ substantially as $\omega$ increases towards   
 $E_F$ before Eq.13 eventually becomes invalid. In the ballistic regime 
the frequency dependence of the effective impedance 
$g(\omega)$ originates from the non-local coupling of the tunneling phase 
$\Phi({\bf 0})$ to the environment
created by the bulk density fluctuations $\rho_{\alpha}({\bf r}\neq {\bf 0})$.

At biases  
$\tau^{-1}<< V<< min\{V_0, E_F\}$ the solution of the renormalization group 
equation (10) 
yields 
$$I(V)\propto  \biggl({V\over E_F}\biggr)^{{1/\nu}}
{1\over |\ln {V/ E_F}|^{2/\nu}}  \eqno(14)$$ 
Unlike disorder scattering,
the long-range Coulomb interactions do affect the $I-V$ characteristic 
even is the limit of vanishing conductivity $\sigma_{xx}$ (see Ref.\cite{11} for a 
comparison), 
although the value of $\sigma_{xx}\propto \tau^{-1}$ \cite{15} determines the 
the lower bound of biases at which Eq.14 applies.
In order to be detectable, the weak $\ln V$ deviation from the power-law $I-V$ 
characteristic 
requires a sizable interval of 
biases, the condition that has not been met in the experiment of 
Ref.\cite{6}.   

To summarize, in the present paper  
we derive a transport theory 
of the QPC between two compressible QH states, each being viewed as a CF metal,  
as well as of the one between a QH state and a normal Fermi-liquid lead.
In our derivation of the zero-dimensional action (1)
we do not rely on the existence of well-defined 1D edges whatsoever.
Nonetheless, the action (1) matches with the QPC theory based upon
the recent refinement of the original phenomenology
of the edges of the incompressible Jain FQHE states, which were shown to allow for a 
single-mode 
description \cite{17}. Moreover, being able to proceed well beyond
the usual limits of the phenomenological approaches, we
consider both cases of short-range (screened) and Coulomb   
interactions, and, respectively, diffusive and ballistic regimes of the bulk 
electron dynamics. 

At small biases, the computed  $I-V$ characteristic is approximately power-law, 
the value of the exponent being in agreement with the available 
experimental data from Ref.\cite{6}. We also predict some departures
from this accustomed power-law behavior which might  
result from the presence of either disorder or long-range Coulomb interactions,
although in the experimentally probed range of 
biases these deviations appear to be small.

\nopagebreak
  
\end{document}